"This manuscript has been authored by UT-Battelle, LLC under Contract No. DE-AC05-00OR22725 with the U.S. Department of Energy. The United States Government retains and the publisher, by accepting the article for publication, acknowledges that the United States Government retains a non-exclusive, paid-up, irrevocable, worldwide license to publish or reproduce the published form of this manuscript, or allow others to do so, for United States Government purposes. The Department of Energy will provide public access to these results of federally sponsored research in accordance with the DOE Public Access Plan (http://energy.gov/downloads/doe-public-access-plan)."

# Oxygen vacancy formation energies in PbTiO$_3$/SrTiO$_3$ superlattice

Lipeng Zhang[1], Isaac Bredeson[1], Axiel Y. Birenbaum[2], P. R. C. Kent[3,4], Valentino R. Cooper[2, §], P. Ganesh[3], Haixuan Xu[1,5]*

[1]Department of Materials Science and Engineering, The University of Tennessee, Knoxville, Tennessee 37996 USA

[2]Materials Science and Technology Division, Oak Ridge National Laboratory, Bethel Valley Road, Oak Ridge, Tennessee 37831, USA

[3]Center for Nanophase Materials Science, Oak Ridge National Laboratory, Bethel Valley Road, Oak Ridge, Tennessee 37831, USA

[4]Computational Science and Engineering Division, Oak Ridge National Laboratory, Bethel Valley Road, Oak Ridge, Tennessee 37831, USA

[5]Joint Institute for Advanced Materials, The University of Tennessee, Knoxville, Tennessee 37996 USA

Email: *xhx@utk.edu; §Coopervr@ornl.gov

## Abstract

The defect stability in a prototypical perovskite oxide superlattice consisting of SrTiO$_3$ and PbTiO$_3$ (STO/PTO) is determined using first principles density functional theory calculations. Specifically, the oxygen vacancy formation energies E$_v$ in the paraelectric and ferroelectric phases of a superlattice with four atomic layers of STO and four layers of PTO (4STO/4PTO) are determined and compared. The effects of charge state, octahedral rotation, polarization, and interfaces on the E$_v$ are examined. The formation energies vary layer-by-layer in the superlattices, with E$_v$ being higher in the ferroelectric phase than that in the paraelectric phase. The two interfaces constructed in these oxide superlattices, which are symmetrically equivalent in the paraelectric systems, exhibit very different formation energies in the ferroelectric superlattices and this can be seen to be driven by the coupling of ferroelectric and rotational modes. At equivalent lattice sites, E$_v$ of charged vacancies is generally lower than that of neutral vacancies. Octahedral rotations (a$^0$a$^0$c$^-$) in the FE superlattice have a significant effect on the E$_v$, increasing the formation energy of vacancies located near the interface but decreasing the formation energy of the oxygen vacancies located in the bulk-like regions of the STO and PTO constituent parts. The formation energy variations among different layers are found to be primarily caused by the difference in the local relaxation at each layer. These fundamental insights into the defect stability in perovskite superlattices can be used to tune defect properties via controlling the constituent materials of superlattices and interface engineering.

## I. Introduction

Oxygen vacancies in perovskite oxides ($ABO_3$) control many physical phenomena, such as ionic conductivity[1-2], catalysis[3-4], and optical properties[5-6]. Understanding of defect stability profiles and their dependence on intrinsic material characteristics is necessary to control the properties of perovskite oxides for practical applications. Extensive research has been published on oxygen vacancy formation energies in bulk $SrTiO_3$[7-10] and other oxides[11-12]. For instance, Deml et al.[13] demonstrated that oxygen vacancy formation energies ($E_v$) are related to the oxide enthalpy of formation, the electronic band gap, and the atomic electronegativity for different oxides.

In oxide superlattices, however, the presence of interfaces between layers of dissimilar oxides can significantly change the defect properties as compared to the bulk oxides[14]. For instance, Zhong et al.[15] showed that $E_v$ varies depending on the proximity to the interface, as well as the thickness of the layers of $SrTiO_3$ and $LaAlO_3$ in the $SrTiO_3/LaAlO_3$ superlattice. In addition, oxide superlattices exhibit various structural distortions, such as octahedral rotation and tilting, which may also influence the defect properties. For example, the ferroelectricity and ground state of $SrTiO_3/PbTiO_3$ (STO/PTO) superlattices are strongly coupled to the rotation of the three-dimensional network of corner-sharing $BO_6$ octahedra across the interface[16-18]. These octahedral rotations can induce changes in electronic and magnetic properties[19-23] and are expected to influence defect stability. Furthermore, Li et al.[16] found that oxygen vacancies can pin the direction of the polarization in both perfect and [001]-oriented superlattices containing oxygen vacancies comprised of alternating single atomic layers of $SrTiO_3/PbTiO_3$ (**1**STO/**1**PTO), indicating the defect properties can also be coupled with polarization. To date, however, the effects of the above-mentioned factors on defect stability and the mutual interaction between interfaces and point defects in oxide superlattices are unclear. It is, therefore, essential to determine the underlying relationship between the intrinsic characteristics of oxide superlattices and oxygen vacancy formation energies, which might be used to control defect properties in these systems for various applications.

In this paper, we examine the effects of polar distortions, octahedral rotations, charge states and interfaces on oxygen vacancy formation energies in **4**STO/**4**PTO superlattices, which is formed by 4 consecutive $PbTiO_3$ layers epitaxially stacked on 4 consecutive $SrTiO_3$ layers in the *z*-direction (Figure 1). This is a prototype perovskite superlattice with complex topologies of ferroelectricity[17, 24-29]. Our goal is to quantify the relevant contributions of these effects in order to extract general trends in this system so that it can serve as a reference for studies of other materials combinations and in studies of vacancy dynamics[30]. We used density functional theory (DFT), which is extensively employed to study defect and physical properties in oxides and oxide superlattices[31-34]. In order to analyze the effect of polarization on the oxygen vacancy $E_v$, we calculated and compared the $E_v$ in the ferroelectric (FE) and paraelectric (PE) phases which possess $a^0a^0c^-$ (in Glazer notation[35]) octahedral rotation. To analyze the effect of charge state and octahedral rotations, charged and neutral $E_v$ were calculated in the paraelectric phase with and without octahedral rotations.

## II. Methodology

The structural and electronic properties of **4**STO/**4**PTO were calculated using DFT via the Vienna *ab initio* simulation package (VASP)[36]. The Perdew-Burke-Ernzerhof functional revised for solids (PBEsol) was used as the exchange-correlation functional. Convergence tests yielded a cutoff energy of 550 eV. The Monkhorst-Pack *k*-point[37] is 2×2×2, which resulted in numerical accuracy to within 1 meV of the total-energy of denser *k*-point meshes. The **4**STO/**4**PTO superlattice relaxations were performed until all Hellmann-Feynman forces were less than 0.01 eV/Å. The valence electron configurations were: Sr 4s4p5s, Pb 5d6s6p, Ti 3p3d4s and O 2s2p. These settings have been shown to yield highly accurate results for predicting $E_v$ in perovskite oxides[8]. For the **4**STO/**4**PTO superlattices, a $2\sqrt{2} \times 2\sqrt{2} \times 8$ supercell was used. This system size was chosen as based on our previous study, to minimize the size effects on the calculated $E_v$ and to balance computational expense[8]. The in-plane lattice parameter was set to 3.898 Å, the optimized value of bulk cubic STO, to mimic the effects of superlattices on a STO substrate[36, 38-39]. In the *z* direction, the *c* lattice parameter was fully relaxed. The STO and PTO reference structures are constructed with and without octahedral rotations with dimensions of $2\sqrt{2} \times 2\sqrt{2} \times 4$ unit cells. The *c/a* of STO and PTO are 1 and 1.015, respectively.

To calculate $E_v$, for each layer, a single oxygen ion was removed and the internal atomic coordinates were fully relaxed, keeping the lattice parameters fixed to the defect-free superlattice. Due to symmetry and the periodicity in the *xy* plane, the formation energies of vacancies on layers with the same *z* coordinate are the same. A single oxygen vacancy formation energy is defined as:

$E_v = E_{tot}(V_O^q) - E_{tot}(\text{ideal}) + \mu + q(E_F + E_{valence} + \Delta V)$

Where $E_v$ is the oxygen vacancy formation energy, $E_{tot}(V_O^q)$ is the total energy of the defective superlattice with one oxygen vacancy in charge state q, in this work q = 2 for a +2 charged oxygen vacancy, and q = 0 for a neutral oxygen vacancy. $E_{tot}(\text{ideal})$ is the total energy of the ideal superlattice, and $\mu$ is the oxygen chemical potential. The chemical potential is set with respect to the equilibrium gas state, $\frac{1}{2}E_{O_2}$, in this case -4.388 eV. The Fermi level, $E_F$, is referenced to the middle of the valence-band maximum and conduction-band minimum of the perfect superlattice. $E_{valence}$, is the valence-band maximum of the ideal superlattice.

The electrostatic potential in the defective superlattice is aligned with that of the perfect one using the correction term $\Delta V$[37].

For comparison and further understanding of structural distortions and their influence on the defect properties, we have performed symmetry mode decompositions on **1**STO/**1**PTO, **2**STO/**2**PTO and **4**STO/**4**PTO for each PE to FE pair using the online tool AMPLIMODES from the Bilbao Crystallographic Server[40]. This allows us to explore the distortions that transform the PE structure to the FE structure expressed as a linear combination of the full basis of mode vectors (in the symmetry-adapted basis), with an amplitude in Angstroms. Hence all distortions can be grouped by their irreducible representations.

**III. Results and Analysis**

a) **Structures of paraelectric and ferroelectric 4PTO/4STO superlattices**

As shown in Table 1, we find that the FE structure has the lowest total energy and can therefore be considered as the ground state[41]. As expected, the *c/a* of the FE phase is larger than that of the PE phase. The atomic structures of these phases are shown in Figure 1.

**Table 1**. The c/a ratio and energy per unit cell relative to the FE phase for the **4**STO/**4**PTO superlattices studied.

| Superlattice structure | Relative energy (meV/unit cell) | c/a |
| --- | --- | --- |
| FE | -- | 1.018 |
| PE | 8 | 1.009 |
| PEWO | 14 | 1.009 |

In Figure 1a the octahedral rotations in the superlattice is shown. In all cases, we observe a Glazer rotation pattern of $a^0a^0c^-$, however, the magnitude of these rotations differs significantly in the PE and FE phases (see Figure 1b and c) In the PE superlattice, the octahedron in the interfacial layer (in purple) has the smallest rotational angle of 5.14°, while the octahedron in the bulk-like regions have rotational angles with magnitudes of 5.51° and 5.25° for STO and PTO, respectively. In the FE phase, however, there are two distinct interfaces due to the polar field within the superlattice. These interfaces have significantly different rotation patterns that depend on the relative orientation of the polarization in the PTO or STO region. If we consider the PTO region, when the polarization points towards the interface there is a large increase in the rotational angle to 7.46°. While the interface at the tail end of the polarization vector has a substantial decrease in the octahedral rotation angle to nearly 0 (0.69°). By symmetry, the opposite is true for the STO region. Furthermore, PTO bulk-like regions (in blue) have larger rotational angles than the STO counterparts (in red). Albeit, in both cases the rotational angle is much smaller than those in PE superlattice phase. These results suggest a strong coupling of the polar modes to the s within the ferroelectric superlattice. The polarization significantly decreases the magnitude of the octahedral rotation on interfaces where the polarization in the PTO region points away from the interface. Meanwhile, when the polarization in the PTO regions points towards the interface it increases the interfacial rotation. In comparison, polarization decreases the rotation angles of the octahedrons locating in the bulk-like region of the FE phase.

To better understand these trends, we perform a symmetry mode decomposition on the **4**STO/**4**PTO superlattices. It is important to note here that our PE structure (Figure 1b) exhibits P4/mbm (127) symmetry which differs slightly to those studied in previous work by Bousquet et al.[17] where they start from a higher symmetry of P4/mmm (123) (see Figure 2a). Our mode decomposition gives a single polar $\Gamma_3^-$ mode with an amplitude of 0.7361 Å. In this case, because of the $2\sqrt{2} \times 2\sqrt{2}$ supercell, the $\Gamma_3^-$ distortion is a superposition of a 1×1 polar (zone center) distortions and M point rotations. In Figure 2b, we see that the cation and anions displace in opposite directions along [001], consistent with an out-of-plane polarization vector.

Furthermore, the oxygen displacements at the interface systematically have [100] & [010] components, indicating the octahedral rotation is coupled to the polar mode. The combined oxygen rotations are constructive at one interface, and destructive at the other. Indeed, the sign of the rotations is the same at both interfaces for mode $M_3^+$, but opposite for $M_1^-$ (see Figure 2b). This explains the interfacial angles measured in the FE structure (Figure 1c): one small (0.69°) and one large (7.46°).

For comparison, we also computed the amplitudes of $\Gamma_3^-$ for shorter period superlattices **1**STO/**1**PTO and **2**STO/**2**PTO. Interestingly, we find that the amplitude of the $\Gamma_3^-$ mode systematically increases from 0.4540 Å in **1**STO/**1**PTO, to 0.7361 Å in **4**STO/**4**PTO; this indicates that polarization should also increase with the number of superlattice layers; in direct agreement with Bousquet et al.'s observation that the FE and AFD displacements are interfacial in origin.

 **(b) Defect formation energies in paraelectric and ferroelectric superlattices**

We first examine the effect of phases (paraelectric vs. ferroelectric) on oxygen vacancy formation energies. In the PE superlattice, layers 1 and 9 are equivalent (the same for layers 3 and 7, etc.) and the $E_v$ are the same in symmetrically equivalent layers, as shown as Figure 3b. The first thing we note is that $E_v$ is higher (by 0.37 eV) in the STO bulk-like region than in the PTO bulk-like region, with the magnitude of $E_v$ in $TiO_2$ layers generally being higher than in neighboring $A$O layers. In $TiO_2$ layers, the lowest charged $E_v$ occurs at the interfaces (layers 1 and 9). In the PTO regions, $E_v$ remains relatively flat, whereas for the STO regions the $E_v$ rises abruptly by roughly 0.5 eV (layer 5 vs. layer 1).

In the FE superlattice, the two interfaces exhibit very different $E_v$ (Figure 3b). In the PTO region, interfaces towards which the polarization points, (layer 1) have significantly higher $E_v$ than those at the tail end (layer 9) of the polarization vector (Figure 3d). (The opposite is true for the STO region.) Based on the symmetry and local polarization analysis, we found $E_v$ strongly depends on polar distortions and octahedral rotations. In addition, the local electrostatic potential along the stacking direction at two interfaces are asymmetric. Although the formation energies of the charged oxygen vacancy in the FE phase are different at two interfaces, the formation energies of a neutral oxygen at two different interfaces are the same. We attribute the formation energy difference between charged oxygen vacancies at the two interfaces to the different redistribution energy of the two positive charges on the vacancy.

Compared with the PE system, there is a substantial increase of $E_v$ for the charged oxygen vacancy shown as Figure 3b. For instance, as depicted in Figure 3c, the $E_v$ in layers 1 and 9 is respectively ~0.9 eV and ~0.3 eV greater in the FE phase as compared to the PE. The formation energy difference between the FE and PE is considered to come from the different structural distortions and variations in the band structures of the two systems. In the bulk-like-layer 5 (STO) and -layer 13 (PTO), the charged oxygen vacancy formation energies are comparable to that in bulk STO and bulk PTO with a difference of 0.02 eV and 0.19 eV, respectively. Again, we see an average 0.49 eV increase of the $E_v$ in the STO layers relative to the PTO layers. Similar to the scenario in the PE system, we find that the $E_v$ in the $TiO_2$ layers are higher than those in neighboring $A$O layers, except for interface layer 9. Also, like the PE superlattices, oxygen vacancies show a preference towards residing in the bulk-like region of PTO (layer 9-12). `

**(c) Effects of Octahedral Rotations**

To understand the role that octahedral rotations play in modulating $E_v$, we compare and contrast $E_v$ in the PE system with and without octahedral rotations (PEWO). As illustrated in Figure 4a and 4c, rotations typically make it less favorable (increases $E_v$) for oxygen vacancies to form, exceptions being oxygen vacancies in bulk-like regions of PTO and STO in the superlattice. Rotations have minimal effects on vacancy formation energies in the AO layer. This correlates well with the observed rotations pattern $a^0a^0c^-$; the octahedral rotations are only around the *z*-axis and would not affect the AO planes. In the case of TiO$_2$ layers, we observe that the formation energies of oxygen vacancies are decreased compared with values without rotation in the bulk-like regions (layers 5 and 13). Conversely, at the interface (layers 1 and 9) octahedral rotations further increases the formation energies of oxygen vacancies.

These results suggest octahedral rotations play different roles on defect stability at different part of the superlattice. At the interfaces, increases in octahedral rotation angles are correlated to increases in E$_v$ (unfavorable). This can be observed in the FE superlattices where the interface with the largest rotation angle (layer 1 with rotation angle of 7.46□) has the largest E$_v$ difference comparing with the PE superlattice. While at layer 9, E$_v$ difference is minimized corresponding to the smaller rotation angle (0.69□). It is worth pointing out that for layer 9 in the FE superlattices the rotation angle is much smaller than the PE phase thus highlighting the importance of coupling of the octahedral rotation and polarization in the superlattices. In the bulk-like regions, octahedral rotations have opposing effects (e.g. rotation decreases E$_v$). Comparing the formation energies of charged oxygen vacancies in reference STO and PTO with and without octahedral rotation, we found that octahedral rotations insignificantly affect the formation energy in the reference STO bulk structure. However, in the reference PTO bulk structure, the octahedral rotation decreases the formation energy by a value of ~0.2 eV for the vacancy on TiO$_2$ layer.

Around the interfaces, octahedral rotation stabilizes the structure and decreases the system energy, resulting in an increase in the formation energies of an oxygen vacancy. In comparison, the bulk oxides (or bulk regions without octahedral rotation) is more stable than the structure with octahedral rotation. Therefore, octahedral rotation increases the potential energy of the bulk-like regions slightly, leading to a reduction of the defect formation energies. In addition, we analyzed the atomistic displacement and atomic strain of these superlattices and the results are supportive of the above analysis. Overall, oxygen vacancies are easier formed at the interface in PEWO than PE superlattice, but are harder formed at the bulk-like region in PEWO than that in the PE superlattice.

**(d) Effects of charge states**

Figure 4a depicts E$_v$ for charged and neutral in PE and PEWO superlattices. Here, it is evident that, the largest effect of charged vacancies is to shift the formation energy down (decrease) relative to the neutral state (see Figure 4b).

In PEWO, the largest difference of the formation energy between the charged and neutral vacancies is observed for the vacancies at the interface, as much as ~2.30 eV; in comparison, the magnitudes of the difference for the vacancies in bulk-like layers 5 and 13, have the smallest values, as low as ~1.8 eV. However, the presence of octahedral rotations, as in the PE phase, changes the profiles for charged and neutral $E_v$ significantly. For instance, the largest difference between the charged and neutral vacancy at the interface decreases to ~1.95 eV, and the difference at layer 5 is no longer the smallest one, as shown in Figure 4b. This further demonstrates that the charge effects on defect formation energies in superlattice systems are coupled with the local octahedral rotation as presented in III (c) and other structural distortions.

**(e) Effect of atomistic relaxation and interface**

A final consideration is to explore the role of atomic rearrangements on the oxygen vacancy formation energies within the superlattice. To do this, the layer-by-layer bond breaking and relaxation energies were calculated for neutral and charged vacancies located in the PEWO and PE superlattices, as shown in Figure 5. The bond breaking energy is here defined as the energy needed to break the bonds between the absent oxygen atom and its nearest bonding atoms. It is calculated by the following formula: $E_{(bonding)} = E_{(static\ defect)} + \frac{1}{2}E_{O2} - E_{(perfect)}$. Where $E_{(static\ defect)}$ is the energy of the un-relaxed structure containing one oxygen vacancy, $\frac{1}{2}E_{O2}$ is the chemical potential of the removed oxygen atom, and $E_{(perfect)}$ is the energy of the structure without the oxygen vacancy. The relaxation energy is the energy change of the superlattice, in which all the ions move from their initial position to the equilibrium position after one oxygen atom removed. It is calculated as: $E_{(relaxation)} = E_{(defect)} - E_{(static\ defect)}$, where $E_{(defect)}$ is the energy of the relaxed structure containing one oxygen vacancy.

For all the investigated scenarios, the bond breaking energy of the vacancies at the interface is 6.33 eV, which is close to the average value (6.32 eV) of the bond breaking energy in STO-bulk and PTO-bulk regions. In addition, the bond breaking energy remains essentially flat for the vacancies in the STO or PTO constituent parts in both PEWO and PE structures, with very small fluctuations of <0.05 eV. The bond breaking energies of the charged and neutral vacancies are the same in PEWO and PE systems, as shown in Figure 5a. This suggests the bond breaking plays a minimal role in creating the variation in the $E_v$ seen in the superlattices with and without octahedral rotations.

On the other hand, the relaxation energies show a strong dependence on their relative distance from the interface. Taking charged $E_v$ in PE as an example, the maximum difference in relaxation energy values in each constituent part are 0.33 eV (STO) and 0.12 eV (PTO), as shown in Figure 5b. These values are comparable with the maximum formation energy difference among the vacancies in STO and PTO constituent parts. The overall differences in $E_v$ as a function of vacancy coordinate are thus related to the relaxation energy rather than bond breaking energy, with the implication that the $E_v$ for a given vacancy position is strongly dependent on the proximity to the interface. Here, phonon mode softening or stiffening may play a significant role in defining how much the lattice may relax around a vacancy site.

In addition, we find that charged vacancies relax more than neutral vacancies, resulting in a lower $E_v$ that is consistent with the previous analysis. Interestingly, we note that the relaxation for charged and neutral oxygen vacancies shows similar trends in PE despite the difference in values. In the PEWO systems, the relaxation energy of the charged and neutral oxygen vacancies also exhibits a similar trend, except for layer 13, in which the relaxation of charged oxygen vacancies is relatively less compared with that in the PE structures.

## IV. Conclusions

In summary, oxygen vacancies show varied formation energies depending on the position of the vacancy along the epitaxial stacking axis. In particular, we find five key trends that seem to correspond to the observed vacancy formation energies: (1) In all cases, $E_v$ is lower in PTO layers than STO. This suggest a preference towards formation of oxygen vacancies in the PTO region of the superlattices. (2) Compared with the neutral oxygen vacancy, charged oxygen vacancies are easier to form in the superlattice, especially for vacancies located at the interface. (3) The charged $E_v$ in the FE phase is higher than that in the PE phase, and the equal values of the $E_v$ for the symmetric position in the PE phase are not equal in the FE phase. (4) Octahedral rotations play a critical role in influence the $E_v$ in FE superlattices. The $a^0a^0c^-$ Glazer rotation pattern minimizes the possible effects of octahedral rotation on $E_v$ in AO layers. (5) The formation of oxygen vacancies couples negatively to polar distortions. This is evident by the fact that the FE phase always has larger $E_v$ than the PE phase, even in cases where there is a large reduction in octahedral rotations.

Overall, octahedral rotations and polar distortions are strongly coupled in the FE superlattice. We find significantly different $E_v$ due to the emergence of different octahedral rotation patterns at the interface. For instance, considering the PTO side of the superlattice if the polar distortions point towards an interface, there is an enhancement in the octahedral rotations which manifests strong enhancements in $E_v$. The opposite is true at the interface at the tail end of the polarization vector. These key trends are manifested through the layer-by-layer relaxation energies which account for the final variations in $E_v$. Ultimately, our work provides insight into the underlying relations between defect properties, polar distortion, octahedral rotation, charge state, and the interfacial effects in perovskite superlattice, which can be used to assist the design and tuning of properties related to the formation of oxygen vacancies in perovskite superlattices.

**Acknowledgements**

This research is sponsored by The University of Tennessee (UT) Science Alliance Joint Directed Research and Development Program (LZ, IB and HX), the Laboratory Directed Research and Development Program of Oak Ridge National Laboratory (VRC, PG and PRCK), managed by UT-Battelle, LLC, for the US Department of Energy (DOE) and the US Department of Energy, Office of Science, Basic Energy Sciences, Materials Sciences and Engineering Division (AYB). This research used resources of The National Institute for Computational Sciences at UT under contract UT-TENN0112 and the National Energy Research Scientific Computing Center, which is supported by the DOE Office of Science under Contract No. DE-AC02-05CH11231

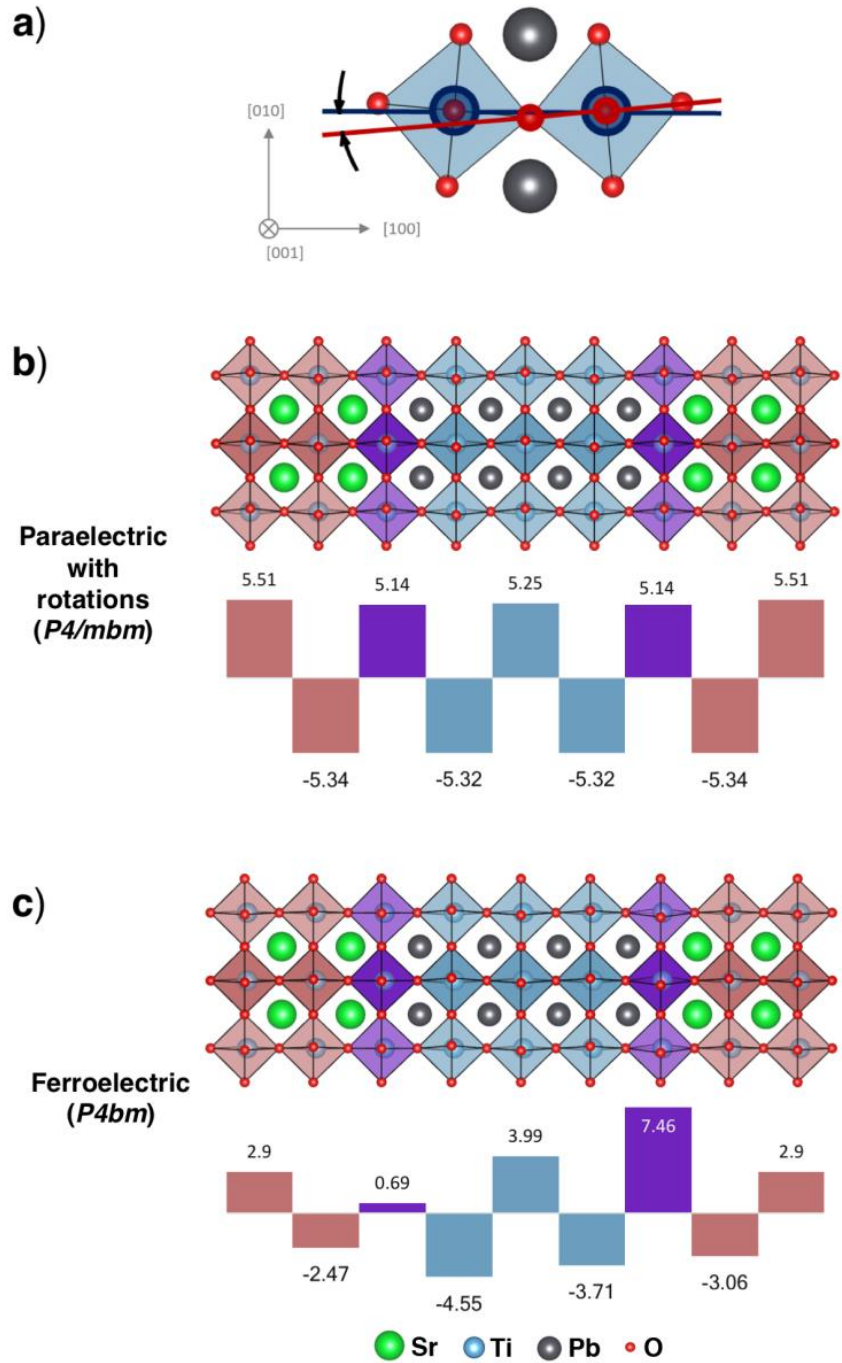

**Figure 1**. (a) The polyhedral rotation is shown with the angle of rotation defined. (b) Paraelectric and (c) ferroelectric **4**STO/**4**PTO superlattices with $a^0a^0c^-$ octahedral rotation. In both cases the graph below denotes the magnitude and direction of the octahedral rotation (about the *z*-axis) for each layer. Purple indicates interfacial layers, while blue and red correspond to PTO and STO layers, respectively. In the FE phase the polarization points along the z-axis towards the right of the figure.

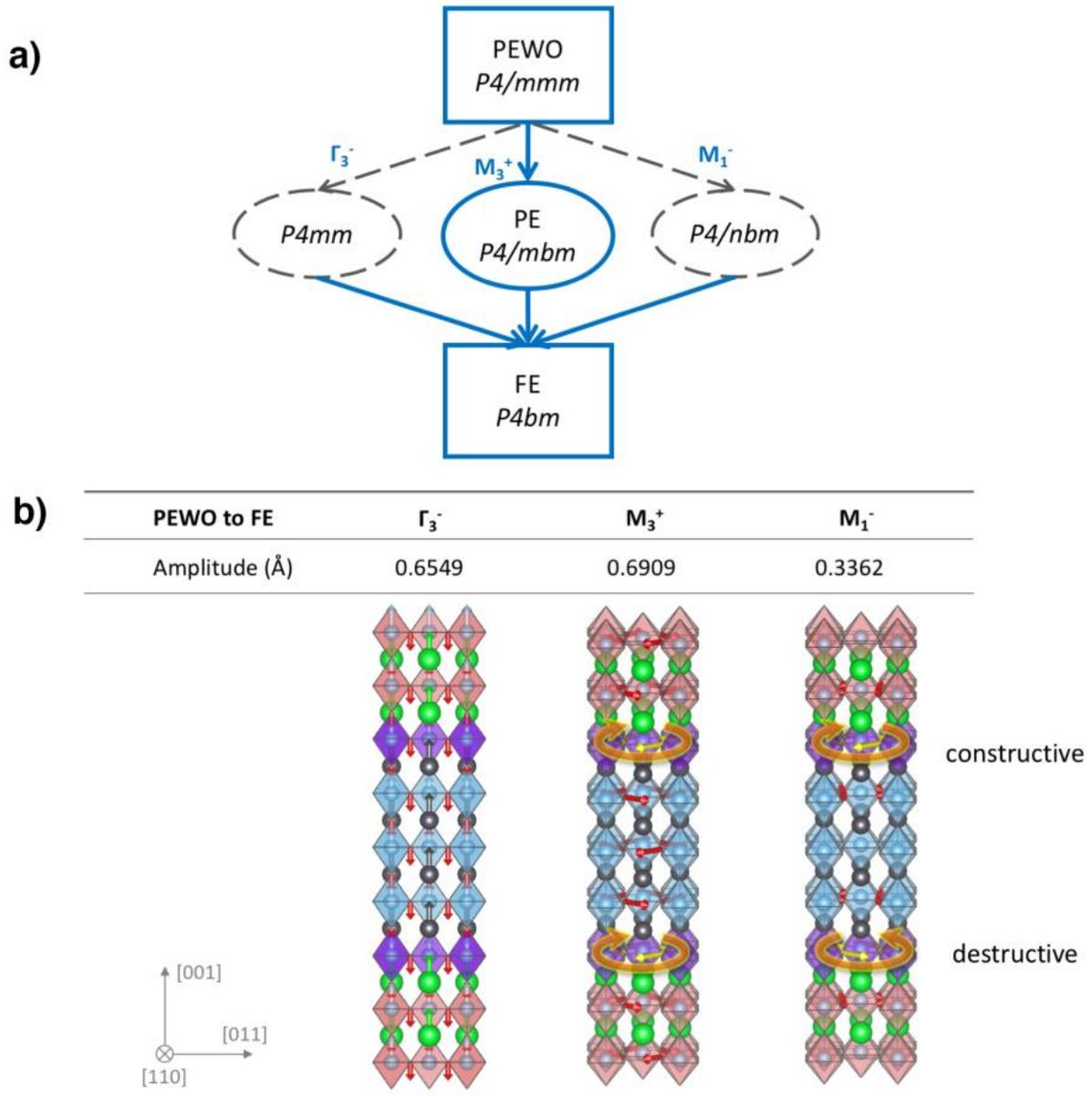

**Figure 2**. Symmetry decomposition on the **4**STO/**4**PTO superlattice. (a) The group-subgroup symmetry tree detailing the intermediate groups between the paraelectric without rotations (PEWO) structure and the ferroelectric FE structure. This indicates that at least two modes are involved in the P4/mmm to P4bm transition. (b) The three distortion modes present when going from the PEWO structure to the FE, with their respective amplitudes and displacement field (arrows). The interfacial oxygen rotations in $M_3^+$ and $M_1^-$ have the same sign at one interface, hence are constructive, and opposite signs at the other interface, hence are destructive.

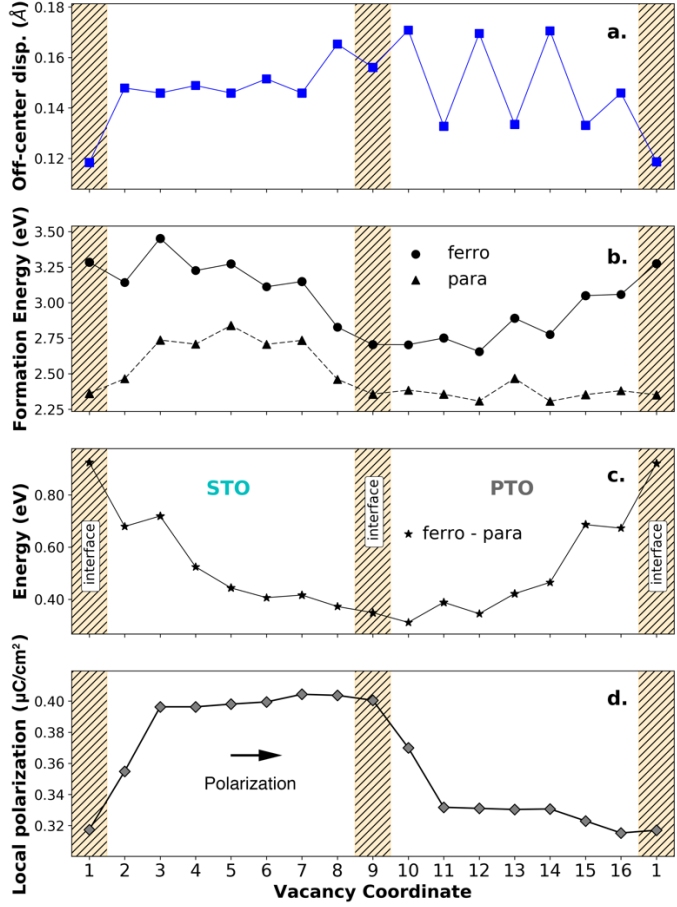

**Figure 3**. (a) Layer-by-layer cation off-center displacements along the *z* direction in the ferroelectric superlattice with octahedral rotations. (b) Formation energies of +2 charged oxygen vacancies in the ferroelectric and paraelectric phases of **4**STO/**4**PTO superlattices. (c) $E_v$ difference between the FE and PE phases for +2 charged vacancies. (d) The layer-by-layer local polarization (out of plane) of the ferroelectric (FE) STO/PTO superlattice. This was calculated with the formula $P_n = \frac{1}{V_n} \sum_i Z_i^* \Delta u_i$, where $P_n$ is local polarization of layer *n*; $Z_i^*$ is the born effective charge of the ion *i* in layer n, $\Delta u_i$ is the off-center displacement in the z direction of ion *i* in the ferroelectric **4**STO/**4**PTO superlattices with octahedral rotations. For the ferroelectric phase the polarization points to the right of the plots. The odd numbered-layers represent $TiO_2$ layers and even numbered layers are AO layers.

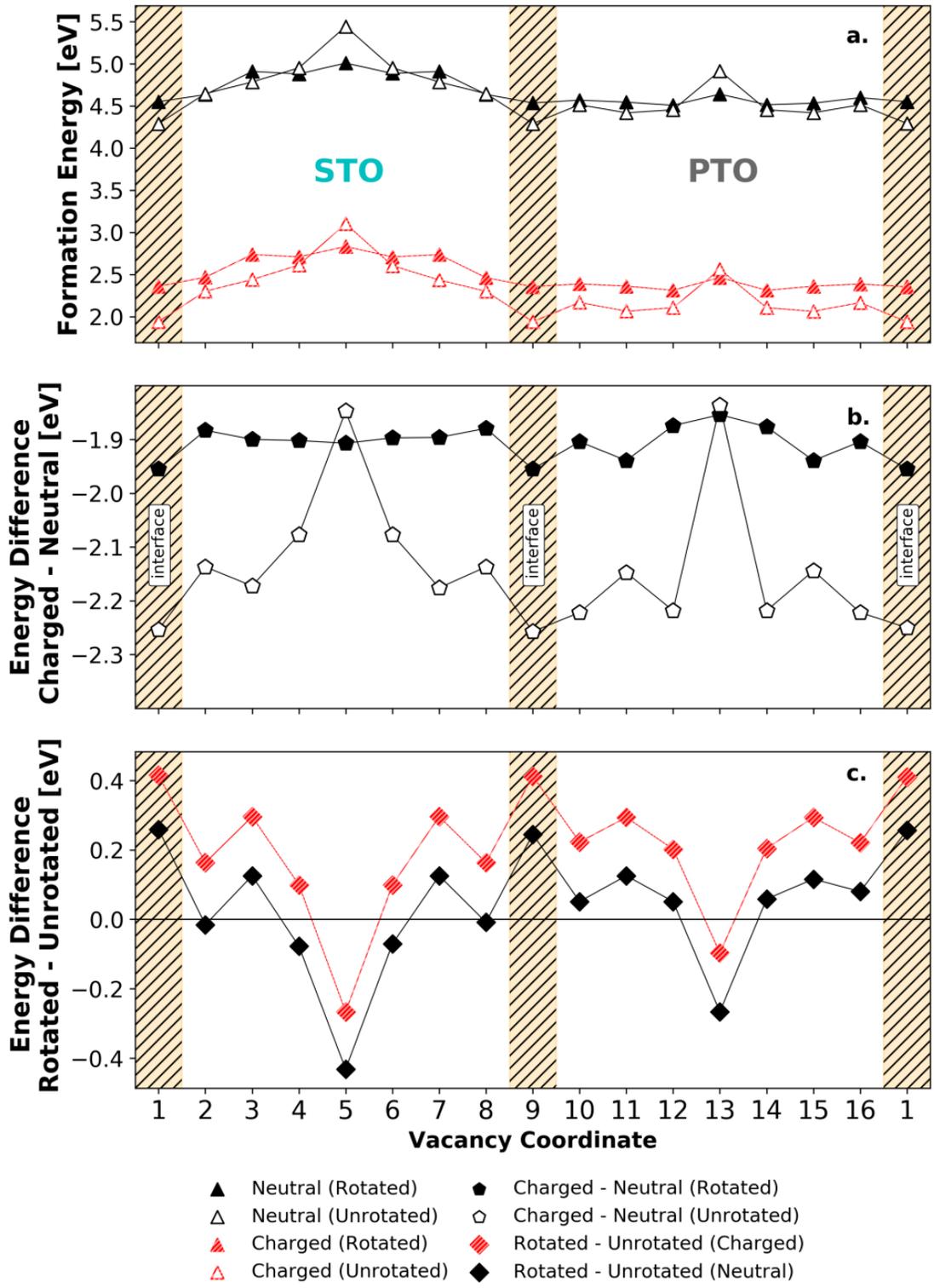

**Figure 4**. (a) The $E_v$ for the paraelectric phase with rotation (PE) and without (PEWO), with charged and neutral vacancies. (b) The charged $E_v$ subtracted by the neutral $E_v$ for PE and PEWO. (c) The rotated (PE) $E_v$ subtracted by the unrotated (PEWO) $E_v$ for charged and neutral vacancies.

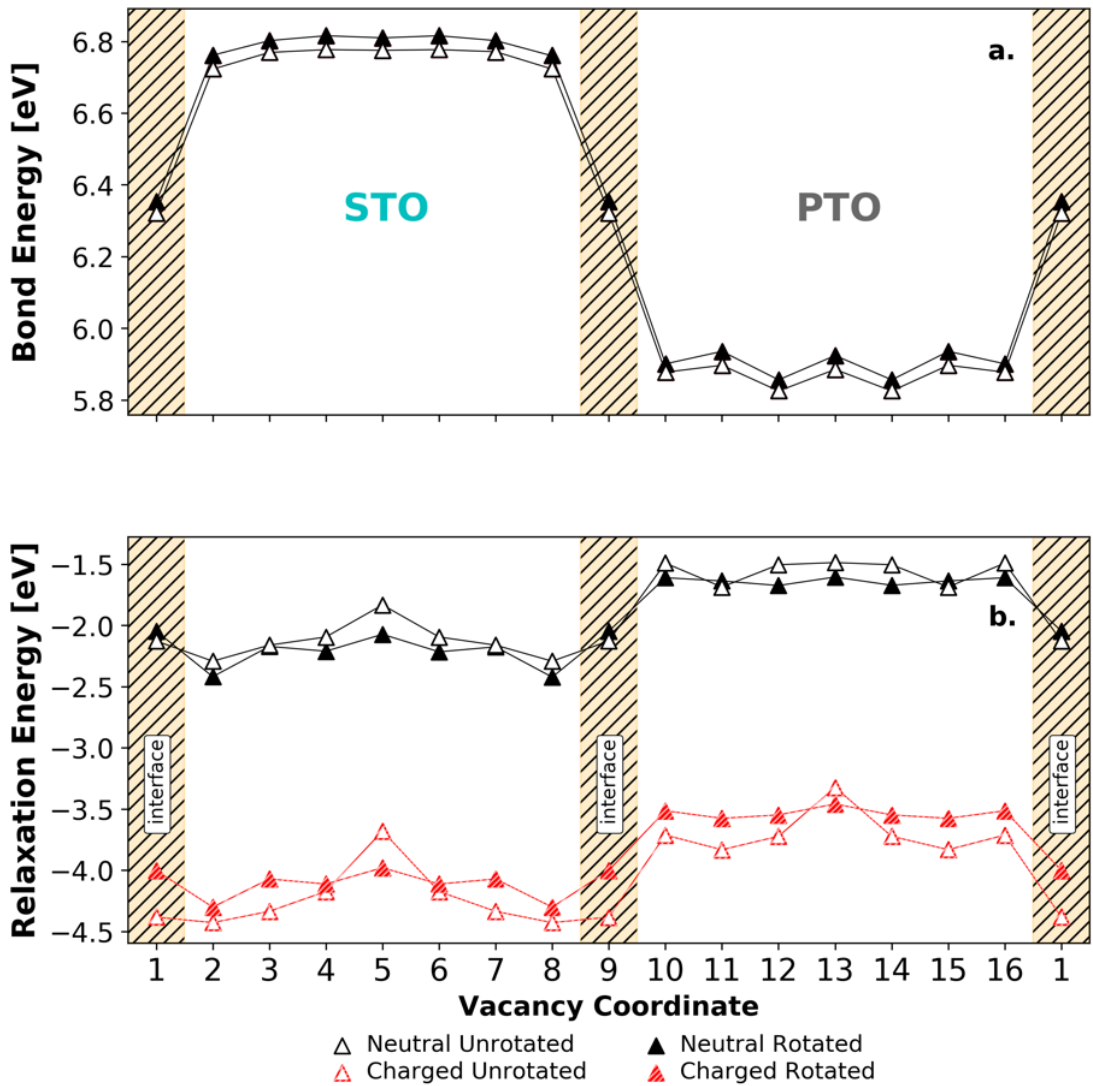

**Figure 5**. The bond energy (a) and relaxation energy (b) for the paraelectric phase with rotation (PE) and without rotation (PEWO), for charged and neutral vacancies, such that $E_v = E_{(bonding)} + E_{(relaxation)}$. In part (a), the unrotated charged and neutral data points overlap, and the rotated charged and neutral overlap, though only the neutral data points are visible.